\documentclass[english]{article}
\usepackage[T1]{fontenc}
\usepackage[latin9]{inputenc}
\usepackage{geometry}
\usepackage{amsmath}
\usepackage{amssymb}
\usepackage{setspace}
\makeatletter
\usepackage{hyperref}
\newcommand{\dd}{\textrm{d}}
\date{}

\makeatother

\usepackage{babel}
\begin{document}
\title{A short note on the appearance of the simplest\\ antilinear ODE in
several physical contexts}
\author{Dmitry Ponomarev$^{1,2,3}$}
\maketitle
\begin{abstract}
In this short note, we review several one-dimensional problems such
as those involving linear Schr{\"o}dinger equation, variable-coefficient
Helmholtz equation, Zakharov-Shabat system and Kubelka-Munk equations.
We show that they all can be reduced to solving one simple antilinear
ordinary differential equation $u^{\prime}\left(x\right)=f\left(x\right)\overline{u\left(x\right)}$
or its nonhomogeneous version $u^{\prime}\left(x\right)=f\left(x\right)\overline{u\left(x\right)}+g\left(x\right)$,
$x\in\left(0,x_{0}\right)\subset\mathbb{R}$. We point out some of
the advantages of the proposed reformulation and call for further
investigation of the obtained ODE.
\end{abstract}

\section{Introduction\label{sec:intro}}

\footnotetext[1]{Institute of Analysis and Scientific Computing, Vienna University of Technology (TU Wien), Austria}\footnotetext[2]{St. Petersburg Department of Steklov Mathematical Institute of Russian Academy of Sciences, Russia}\footnotetext[3]{Contact: dmitry.ponomarev@asc.tuwien.ac.at}
Many physical phenomena can be directly described by or reduced to
systems of differential equations having certain structural properties.
Restricting ourselves here to linear one-dimensional settings, we
are concerned with a pair of first-order ODEs whose matrix is antidiagonal
with complex-conjugate elements. Namely, given $x_{0}\in\mathbb{R}$
and complex-valued function $f\left(x\right)$, we consider the equation
\begin{equation}
U^{\prime}\left(x\right)=\left(\begin{array}{cc}
0 & f\left(x\right)\\
\overline{f\left(x\right)} & 0
\end{array}\right)U\left(x\right),\hspace{1em}x\in\left(0,x_{0}\right),\label{eq:U_gen_antidiag}
\end{equation}
as well as its nonhomogeneous analog
\begin{equation}
U^{\prime}\left(x\right)=\left(\begin{array}{cc}
0 & f\left(x\right)\\
\overline{f\left(x\right)} & 0
\end{array}\right)U\left(x\right)+G\left(x\right),\hspace{1em}x\in\left(0,x_{0}\right)\text{,}\label{eq:U_gen_antidiag_inhom}
\end{equation}
where $U\left(x\right)\equiv\left(u_{1}\left(x\right),u_{2}\left(x\right)\right)^{T}\in\mathbb{C}^{2}$
is an unknown solution-vector, $G\left(x\right)\equiv\left(g_{1}\left(x\right),g_{2}\left(x\right)\right)^{T}\in\mathbb{C}^{2}$
is a given vector-function, and each of equations (\ref{eq:U_gen_antidiag})--(\ref{eq:U_gen_antidiag_inhom})
is supplemented by the initial condition $U\left(0\right)=U_{0}\in\mathbb{C}^{2}$.
Here and onwards, we employ the notation $\overline{\cdot}$ to denote
complex conjugation.

Similarly to Hamiltonian, Dirac and more general canonical systems
(see e.g. \cite{Remling}), equations (\ref{eq:U_gen_antidiag})--(\ref{eq:U_gen_antidiag_inhom})
constitute an important class of dynamical systems for two reasons.
On the one hand, as we shall further see, formulations of several
important problems are reducible to either (\ref{eq:U_gen_antidiag})
or (\ref{eq:U_gen_antidiag_inhom}). On the other hand, these systems
are close to being exactly solvable in the following sense. Let us
focus on (\ref{eq:U_gen_antidiag}) and consider the more general
system
\begin{equation}
U^{\prime}\left(x\right)=\left(\begin{array}{cc}
p\left(x\right) & r\left(x\right)\\
s\left(x\right) & q\left(x\right)
\end{array}\right)U\left(x\right),\hspace{1em}x\in\left(0,x_{0}\right).\label{eq:U_pqrs_eq}
\end{equation}
We note that the diagonal elements in the matrix of (\ref{eq:U_pqrs_eq})
can be removed by the exponential multiplier transform. Namely, by
setting 
\[
V\left(x\right):=\left(\begin{array}{cc}
e^{-\int_{0}^{x}p\left(\tau\right)\dd\tau} & 0\\
0 & e^{-\int_{0}^{x}q\left(\tau\right)\dd\tau}
\end{array}\right)U\left(x\right),
\]
one can observe that $V\left(x\right)$ satisfies
\begin{equation}
V^{\prime}\left(x\right)=\left(\begin{array}{cc}
0 & r\left(x\right)\exp\left(-\int_{0}^{x}\left[p\left(\tau\right)-q\left(\tau\right)\right]\dd\tau\right)\\
s\left(x\right)\exp\left(\int_{0}^{x}\left[p\left(\tau\right)-q\left(\tau\right)\right]\dd\tau\right) & 0
\end{array}\right)V\left(x\right),\hspace{1em}x\in\left(0,x_{0}\right),\label{eq:V_gen_antidiag}
\end{equation}
with the initial condition $V\left(0\right)=U_{0}$. Now, if the anti-diagonal
elements of the matrix in the right-hand side of (\ref{eq:V_gen_antidiag})
are equal, i.e.
\begin{equation}
r\left(x\right)\exp\left(-\int_{0}^{x}\left[p\left(\tau\right)-q\left(\tau\right)\right]\dd\tau\right)=s\left(x\right)\exp\left(\int_{0}^{x}\left[p\left(\tau\right)-q\left(\tau\right)\right]\dd\tau\right)=:c_{1}\left(x\right),\label{eq:antidiag_cond}
\end{equation}
then the solution can be written explicitly as
\[
V\left(x\right)=\left[\cosh\left(\int_{0}^{x}c_{1}\left(\tau\right)\dd\tau\right)I+\sinh\left(\int_{0}^{x}c_{1}\left(\tau\right)\dd\tau\right)S\right]U_{0},
\]
where
\begin{equation}
I:=\left(\begin{array}{cc}
1 & 0\\
0 & 1
\end{array}\right),\hspace{1em}\hspace{1em}S:=\left(\begin{array}{cc}
0 & 1\\
1 & 0
\end{array}\right).\label{eq:I_S_def}
\end{equation}
However, condition (\ref{eq:antidiag_cond}), which amounts to the
assumption
\begin{equation}
s\left(x\right)=r\left(x\right)\exp\left(-2\int_{0}^{x}\left[p\left(\tau\right)-q\left(\tau\right)\right]\dd\tau\right),\label{eq:sr_cond_strong}
\end{equation}
may be too restrictive. Indeed, in view of multiple possible similarity
transformations allowing to rewrite (\ref{eq:U_pqrs_eq}) in different
equivalent forms, we want to have a clearly identifiable matrix structure
which should be, on the one hand, immediately recognisable and, on
the other hand, leading to a solution simplification or even an explicit
solution. Such an identifiable structure may be, for example, a pairwise
relation between some of the elements of the matrix in (\ref{eq:U_pqrs_eq}).
The explicit solvability condition, nevertheless, plays against any
visible structural property of the matrix: even though assumption
(\ref{eq:sr_cond_strong}) leads to a closed-form solution, it implies
a rather complicated relation between the matrix elements. Condition
(\ref{eq:sr_cond_strong}) is very specific and thus unlikely to be
satisfied for any easily describable class of matrix elements unless,
of course, $p\equiv q$, which would then also entail that $r\equiv s$.
This exactly solvable case with equal diagonal and anti-diagonal elements
in (\ref{eq:U_pqrs_eq}) may be sometimes valuable but it does not
seem to be the one that covers many fruitful applications.

It turns out that condition (\ref{eq:sr_cond_strong}) has an analog
which is less stringent in form of the matrix elements in (\ref{eq:U_pqrs_eq}),
with more pertinence to important physical contexts, and, at the same
time, it still leads to a significant simplification of the solution
procedure (and, at least in some cases, also to closed-form solutions).
This condition reads 
\begin{equation}
s\left(x\right)=\overline{r\left(x\right)}\exp\left(-2\text{ Re}\int_{0}^{x}\left[p\left(\tau\right)-q\left(\tau\right)\right]\dd\tau\right).\label{eq:sr_cond_weak}
\end{equation}
Despite the similarity to (\ref{eq:sr_cond_strong}), condition (\ref{eq:sr_cond_weak})
is easier to satisfy while preserving a visible matrix structure.
Indeed, if $p\left(x\right)-q\left(x\right)$ is a purely imaginary
function (e.g., in particular, when $p\equiv\overline{q}$), the complicated
exponential factor in (\ref{eq:sr_cond_weak}) disappears. In this
case, the implied condition $s\left(x\right)=\overline{r\left(x\right)}$
is clearly indentifiable but far from being trivial since, as we shall
see, it covers a variety of different practical applications. This
reasoning motivates us to consider (\ref{eq:U_gen_antidiag}) as well
as its nonhogeneous analog (\ref{eq:U_gen_antidiag_inhom}).

The plan of this note is as follows. Section \ref{sec:transform}
is dedicated to the transformation of (\ref{eq:U_gen_antidiag}) and
(\ref{eq:U_gen_antidiag_inhom}) into the homogeneous antilinear ODE
\begin{equation}
u^{\prime}\left(x\right)=f\left(x\right)\overline{u\left(x\right)},\hspace{1em}x\in\left(0,x_{0}\right),\label{eq:u_antilin_ODE}
\end{equation}
and its nonhomogeneous analog 
\begin{equation}
u^{\prime}\left(x\right)=f\left(x\right)\overline{u\left(x\right)}+g\left(x\right),\hspace{1em}x\in\left(0,x_{0}\right),\label{eq:u_antilin_ODE_inhom}
\end{equation}
respectively. Next, in Section \ref{sec:contexts}, we outline some
relevant physical applications, i.e. problems which, upon appropriate
transformations, can be recast as (\ref{eq:U_gen_antidiag}) or (\ref{eq:U_gen_antidiag_inhom})
and are thus reducible to formulations involving antilinear ODE (\ref{eq:u_antilin_ODE}),
or, in one case, its nonhomogeneous version (\ref{eq:u_antilin_ODE_inhom}).
Finally, in Section \ref{sec:concl}, we conclude with some remarks
on how antilinear ODEs can be constructively addressed further and
briefly mention a couple of other applications.

\section{Transformation of an antidiagonal problem into an antilinear ODE\label{sec:transform}}

\subsection{Homogeneous case: from (\ref{eq:U_gen_antidiag}) to (\ref{eq:u_antilin_ODE})\label{subsec:transform_hom}}

We consider (\ref{eq:U_gen_antidiag}) supplemented with the initial
data $U\left(0\right)=U_{0}\equiv\left(u_{1}^{0},u_{2}^{0}\right)^{T}$,
where $u_{1}^{0}$, $u_{2}^{0}\in\mathbb{C}$ are arbitrary constants,
and we devise a transformation that allows construction of the solution
of system (\ref{eq:U_gen_antidiag}) in terms of the solution of an
antilinear ODE of the form (\ref{eq:u_antilin_ODE}).

Let us first motivate our approach to construction of such a transformation.
To this effect, given a complex-valued function $f\left(x\right)$,
it is instructive to consider the elementary differential equation
for $v\left(x\right)$
\begin{equation}
v^{\prime}\left(x\right)=f\left(x\right)v\left(x\right),\hspace{1em}x\in\left(0,x_{0}\right),\label{eq:v_ODE}
\end{equation}
with the initial condition $v\left(0\right)=1$. On the one hand,
(\ref{eq:v_ODE}) is in separable form and hence can be integrated
directly to yield the solution
\begin{equation}
v\left(x\right)=\exp\left(\int_{0}^{x}f\left(\tau\right)\dd\tau\right).\label{eq:v_exp_sol}
\end{equation}
On the other hand, rewriting (\ref{eq:v_ODE}) in the integral form
\[
v\left(x\right)=1+\int_{0}^{x}f\left(\tau\right)v\left(\tau\right)\dd\tau,\hspace{1em}x\in\left(0,x_{0}\right),
\]
the Picard iterative process gives
\begin{equation}
v\left(x\right)=1+\int_{0}^{x}f\left(\tau\right)\dd\tau+\int_{0}^{x}f\left(\tau_{2}\right)\int_{0}^{\tau_{2}}f\left(\tau_{1}\right)\dd\tau_{1}\dd\tau_{2}+\int_{0}^{x}f\left(\tau_{3}\right)\int_{0}^{\tau_{3}}f\left(\tau_{2}\right)\int_{0}^{\tau_{2}}f\left(\tau_{1}\right)\dd\tau_{1}\dd\tau_{2}\dd\tau_{3}+\ldots.\label{eq:v_Pic_sol}
\end{equation}
Comparison of (\ref{eq:v_exp_sol}) with (\ref{eq:v_Pic_sol}) results
in important identities
\[
\exp\left(\int_{0}^{x}f\left(\tau\right)\dd\tau\right)=1+\int_{0}^{x}f\left(\tau\right)\dd\tau+\int_{0}^{x}f\left(\tau_{2}\right)\int_{0}^{\tau_{2}}f\left(\tau_{1}\right)\dd\tau_{1}\dd\tau_{2}+\int_{0}^{x}f\left(\tau_{3}\right)\int_{0}^{\tau_{3}}f\left(\tau_{2}\right)\int_{0}^{\tau_{2}}f\left(\tau_{1}\right)\dd\tau_{1}\dd\tau_{2}\dd\tau_{3}+\ldots,
\]
\begin{equation}
\sinh\left(\int_{0}^{x}f\left(\tau\right)\dd\tau\right)=\int_{0}^{x}f\left(\tau\right)\dd\tau+\int_{0}^{x}f\left(\tau_{3}\right)\int_{0}^{\tau_{3}}f\left(\tau_{2}\right)\int_{0}^{\tau_{2}}f\left(\tau_{1}\right)\dd\tau_{1}\dd\tau_{2}\dd\tau_{3}+\ldots,\label{eq:sinh_id}
\end{equation}
\begin{equation}
\cosh\left(\int_{0}^{x}f\left(\tau\right)\dd\tau\right)=1+\int_{0}^{x}f\left(\tau_{2}\right)\int_{0}^{\tau_{2}}f\left(\tau_{1}\right)\dd\tau_{1}\dd\tau_{2}+\int_{0}^{x}f\left(\tau_{4}\right)\int_{0}^{\tau_{4}}f\left(\tau_{3}\right)\int_{0}^{\tau_{3}}f\left(\tau_{2}\right)\int_{0}^{\tau_{2}}f\left(\tau_{1}\right)\dd\tau_{1}\dd\tau_{2}\dd\tau_{3}\dd\tau_{4}+\ldots,\label{eq:cosh_id}
\end{equation}
where we used the identity $\exp\left(z\right)=\cosh\left(z\right)+\sinh\left(z\right)$,
$z\in\mathbb{C}$, and the parity argument to split the terms: $\sinh$
is an odd function and hence (\ref{eq:sinh_id}) may contain only
odd number of multiplicative instances of $f$, and similarly, (\ref{eq:cosh_id})
may contain only terms with even number of multiplications by $f$
due to $\cosh$ being an even function. 

Now, similarly to (\ref{eq:sinh_id})--(\ref{eq:cosh_id}), let us
consider the following quantities
\begin{equation}
S_{f}\left(x\right):=\int_{0}^{x}f\left(\tau_{1}\right)\dd\tau_{1}+\int_{0}^{x}f\left(\tau_{3}\right)\int_{0}^{\tau_{3}}\overline{f\left(\tau_{2}\right)}\int_{0}^{\tau_{2}}f\left(\tau_{1}\right)\dd\tau_{1}\dd\tau_{2}\dd\tau_{3}+\ldots,\label{eq:Sf_def}
\end{equation}
\begin{equation}
C_{f}\left(x\right):=1+\int_{0}^{x}f\left(\tau_{2}\right)\int_{0}^{\tau_{2}}\overline{f\left(\tau_{1}\right)}\dd\tau_{1}\dd\tau_{2}+\int_{0}^{x}f\left(\tau_{4}\right)\int_{0}^{\tau_{4}}\overline{f\left(\tau_{3}\right)}\int_{0}^{\tau_{3}}f\left(\tau_{2}\right)\int_{0}^{\tau_{2}}\overline{f\left(\tau_{1}\right)}\dd\tau_{1}\dd\tau_{2}\dd\tau_{3}\dd\tau_{4}+\ldots.\label{eq:Cf_def}
\end{equation}
Let us show that (\ref{eq:Sf_def})--(\ref{eq:Cf_def}) are inherent
to an algebraic structure underlying (\ref{eq:U_gen_antidiag}). To
this effect, we rewrite (\ref{eq:U_gen_antidiag}) in the integral
form
\begin{equation}
U\left(x\right)=U_{0}+\int_{0}^{x}A\left(\tau\right)U\left(\tau\right)\dd\tau,\hspace{1em}x\in\left(0,x_{0}\right),\hspace{1em}\hspace{1em}A\left(\tau\right):=\left(\begin{array}{cc}
0 & f\left(\tau\right)\\
\overline{f\left(\tau\right)} & 0
\end{array}\right),\label{eq:U_int_form}
\end{equation}
and note that
\[
A\left(\tau_{2}\right)A\left(\tau_{1}\right)=\left(\begin{array}{cc}
f\left(\tau_{2}\right)\overline{f\left(\tau_{1}\right)} & 0\\
0 & \overline{f\left(\tau_{2}\right)}f\left(\tau_{1}\right)
\end{array}\right),\hspace{1em}A\left(\tau_{3}\right)A\left(\tau_{2}\right)A\left(\tau_{1}\right)=\left(\begin{array}{cc}
0 & f\left(\tau_{3}\right)\overline{f\left(\tau_{2}\right)}f\left(\tau_{1}\right)\\
\overline{f\left(\tau_{3}\right)}f\left(\tau_{2}\right)\overline{f\left(\tau_{1}\right)} & 0
\end{array}\right),
\]
\[
A\left(\tau_{4}\right)A\left(\tau_{3}\right)A\left(\tau_{2}\right)A\left(\tau_{1}\right)=\left(\begin{array}{cc}
f\left(\tau_{4}\right)\overline{f\left(\tau_{3}\right)}f\left(\tau_{2}\right)\overline{f\left(\tau_{1}\right)} & 0\\
0 & \overline{f\left(\tau_{4}\right)}f\left(\tau_{3}\right)\overline{f\left(\tau_{2}\right)}f\left(\tau_{1}\right)
\end{array}\right),\hspace{1em}\ldots.
\]
Therefore, writing out Picard iterations for solving (\ref{eq:U_int_form}),
we obtain
\begin{equation}
U\left(x\right)=\left(\begin{array}{cc}
C_{f}\left(x\right) & S_{f}\left(x\right)\\
S_{\overline{f}}\left(x\right) & C_{\overline{f}}\left(x\right)
\end{array}\right)U_{0}=\left(\begin{array}{cc}
C_{f}\left(x\right) & S_{f}\left(x\right)\\
\overline{S_{f}\left(x\right)} & \overline{C_{f}\left(x\right)}
\end{array}\right)U_{0}.\label{eq:U_sol}
\end{equation}
Furthermore, it is easy to see from (\ref{eq:Sf_def})--(\ref{eq:Cf_def})
that $S_{f}\left(x\right)$, $C_{f}\left(x\right)$ obey the following
intertwining relation
\begin{equation}
C_{f}^{\prime}\left(x\right)=f\left(x\right)\overline{S_{f}\left(x\right)},\hspace{1em}\hspace{1em}S_{f}^{\prime}\left(x\right)=f\left(x\right)\overline{C_{f}\left(x\right)},\hspace{1em}\hspace{1em}x\in\left(0,x_{0}\right),\label{eq:Sf_Cf_coupl}
\end{equation}
and the conditions $S_{f}\left(0\right)=0$, $C_{f}\left(0\right)=1$.
Introducing another pair of functions 
\[
Z_{+}\left(x\right):=C_{f}\left(x\right)+S_{f}\left(x\right),\hspace{1em}\hspace{1em}Z_{-}\left(x\right):=C_{f}\left(x\right)-S_{f}\left(x\right),
\]
we decouple (\ref{eq:Sf_Cf_coupl}) as
\begin{equation}
Z_{+}^{\prime}\left(x\right)=f\left(x\right)\overline{Z_{+}\left(x\right)},\hspace{1em}x\in\left(0,x_{0}\right),\hspace{1em}\hspace{1em}Z_{+}\left(0\right)=1,\label{eq:Z_pl_ODE}
\end{equation}
\begin{equation}
Z_{-}^{\prime}\left(x\right)=-f\left(x\right)\overline{Z_{-}\left(x\right)},\hspace{1em}x\in\left(0,x_{0}\right),\hspace{1em}\hspace{1em}Z_{-}\left(0\right)=1.\label{eq:Z_min_ODE}
\end{equation}
Equations (\ref{eq:Z_pl_ODE})--(\ref{eq:Z_min_ODE}) are two separate
instances of the initial-value problem featuring the antilinear ODE
given by (\ref{eq:u_antilin_ODE}). Solution of this ODE would thus
yield the solutions of (\ref{eq:Z_pl_ODE})--(\ref{eq:Z_min_ODE})
and, consequently, also of (\ref{eq:Sf_Cf_coupl}), providing $S_{f}\left(x\right)$,
$C_{f}\left(x\right)$ appearing in (\ref{eq:U_sol}) which furnishes
the solution of (\ref{eq:U_gen_antidiag}).

\subsection{Nonhomogeneous case: from (\ref{eq:U_gen_antidiag_inhom}) to (\ref{eq:u_antilin_ODE_inhom})\label{subsec:transform_inhom}}

Let us now consider (\ref{eq:U_gen_antidiag_inhom}) with $G\left(x\right)\equiv\left(g_{1}\left(x\right),g_{2}\left(x\right)\right)^{T}$
and subject to the initial condition $U\left(0\right)=U_{0}\equiv\left(u_{1}^{0},u_{2}^{0}\right)^{T}$.
We are going to show that, in particular case where
\begin{equation}
g_{2}\left(x\right)=i\,\overline{g_{1}\left(x\right)},\hspace{1em}\hspace{1em}u_{2}^{0}=i\,\overline{u_{1}^{0}},\label{eq:f_U0_cond}
\end{equation}
the solution of (\ref{eq:U_gen_antidiag_inhom}) can be constucted
in terms of solutions of two instances of problem (\ref{eq:u_antilin_ODE_inhom}).
As we shall see in Subsection \ref{subsec:cont_Helmh}, assumption
(\ref{eq:f_U0_cond}) will be satisfied in at least one important
practical context.

Similarly to (\ref{eq:Sf_def})--(\ref{eq:Cf_def}), let us introduce
\begin{equation}
\mathrm{S}_{f,h}\left(x\right):=\int_{0}^{x}f\left(\tau_{1}\right)h\left(\tau_{1}\right)\dd\tau_{1}+\int_{0}^{x}f\left(\tau_{3}\right)\int_{0}^{\tau_{3}}\overline{f\left(\tau_{2}\right)}\int_{0}^{\tau_{2}}f\left(\tau_{1}\right)h\left(\tau_{1}\right)\dd\tau_{1}\dd\tau_{2}\dd\tau_{3}+\ldots,\label{eq:Sfh_def}
\end{equation}
\begin{align}
\mathrm{C}_{f,h}\left(x\right):= & h\left(x\right)+\int_{0}^{x}f\left(\tau_{2}\right)\int_{0}^{\tau_{2}}\overline{f\left(\tau_{1}\right)}h\left(\tau_{1}\right)\dd\tau_{1}\dd\tau_{2}\label{eq:Cfh_def}\\
 & +\int_{0}^{x}f\left(\tau_{4}\right)\int_{0}^{\tau_{4}}\overline{f\left(\tau_{3}\right)}\int_{0}^{\tau_{3}}f\left(\tau_{2}\right)\int_{0}^{\tau_{2}}\overline{f\left(\tau_{1}\right)}h\left(\tau_{1}\right)\dd\tau_{1}\dd\tau_{2}\dd\tau_{3}\dd\tau_{4}+\ldots.\nonumber 
\end{align}
Rewriting (\ref{eq:U_gen_antidiag_inhom}) in the integral form

\begin{equation}
U\left(x\right)=U_{0}+\int_{0}^{x}G\left(\tau\right)\dd\tau+\int_{0}^{x}A\left(\tau\right)U\left(\tau\right)\dd\tau,\hspace{1em}x\in\left(0,x_{0}\right),\hspace{1em}\hspace{1em}A\left(\tau\right):=\left(\begin{array}{cc}
0 & f\left(\tau\right)\\
\overline{f\left(\tau\right)} & 0
\end{array}\right),\label{eq:U_int_form_inhom}
\end{equation}
it is straightforward to see that Picard iterations give
\[
U\left(x\right)=\left(\begin{array}{c}
\mathrm{S}_{f,h_{2}}\left(x\right)+\mathrm{C}_{f,h_{1}}\left(x\right)\\
\mathrm{S}_{\overline{f},h_{1}}\left(x\right)+\mathrm{C}_{\overline{f},h_{2}}\left(x\right)
\end{array}\right),
\]
where $\mathrm{S}_{f,h}\left(x\right)$, $\mathrm{C}_{f,h}\left(x\right)$
are as defined by (\ref{eq:Sfh_def})--(\ref{eq:Cfh_def}), and
\begin{equation}
h_{1}\left(x\right):=u_{1}^{0}+\int_{0}^{x}g_{1}\left(\tau\right)\dd\tau,\hspace{1em}\hspace{1em}h_{2}\left(x\right):=u_{2}^{0}+\int_{0}^{x}g_{2}\left(\tau\right)\dd\tau.\label{eq:h1h2_def}
\end{equation}
By means of differentiation of $\mathrm{S}_{f,h_{2}}\left(x\right)$
and $\mathrm{C}_{f,h_{1}}\left(x\right)$, we obtain the following
intertwining relation
\begin{equation}
\mathrm{C}_{f,h_{1}}^{\prime}\left(x\right)=h_{1}^{\prime}\left(x\right)+f\left(x\right)\mathrm{S}_{\overline{f},h_{1}}\left(x\right),\hspace{1em}\mathrm{S}_{f,h_{2}}^{\prime}\left(x\right)=f\left(x\right)\mathrm{C}_{\overline{f},h_{2}}\left(x\right),\hspace{1em}\hspace{1em}x\in\left(0,x_{0}\right),\label{eq:Cfh_Sfh_coupl}
\end{equation}
which is to be supplemented by the conditions $\mathrm{C}_{f,h_{1}}\left(0\right)=h_{1}\left(0\right)$,
$\mathrm{S}_{f,h_{2}}\left(0\right)=0$.

Note that, from (\ref{eq:Sfh_def})--(\ref{eq:Cfh_def}), $\mathrm{S}_{\overline{f},h_{1}}\left(x\right)=\overline{\mathrm{S}_{f,\overline{h_{1}}}\left(x\right)}$
and $\mathrm{C}_{\overline{f},h_{2}}\left(x\right)=\overline{\mathrm{C}_{f,\overline{h_{2}}}\left(x\right)}$.
Moreover, assumption (\ref{eq:f_U0_cond}) entails that $h_{2}\left(x\right)=i\,\overline{h_{1}\left(x\right)}$,
and by linearity in $h$ of $\mathrm{S}_{f,h}\left(x\right)$, $\mathrm{C}_{f,h}\left(x\right)$,
we have $\mathrm{S}_{f,h_{2}}\left(x\right)=i\,\mathrm{S}_{f,\overline{h_{1}}}\left(x\right)$
and $\overline{\mathrm{C}_{f,\overline{h_{2}}}\left(x\right)}=i\,\overline{\mathrm{C}_{f,h_{1}}\left(x\right)}$.
Consequently, relation (\ref{eq:Cfh_Sfh_coupl}) becomes 
\begin{equation}
\mathrm{C}_{f,h_{1}}^{\prime}\left(x\right)=h_{1}^{\prime}\left(x\right)+f\left(x\right)\overline{\mathrm{S}_{f,\overline{h_{1}}}\left(x\right)},\hspace{1em}\mathrm{S}_{f,\overline{h_{1}}}^{\prime}\left(x\right)=f\left(x\right)\overline{\mathrm{C}_{f,h_{1}}\left(x\right)},\hspace{1em}\hspace{1em}x\in\left(0,x_{0}\right).\label{eq:Cfh_Sfh_coupl_simpl}
\end{equation}
Setting 
\begin{equation}
Z_{+}\left(x\right):=\mathrm{C}_{f,h_{1}}\left(x\right)+\mathrm{S}_{f,\overline{h_{1}}}\left(x\right),\hspace{1em}Z_{-}\left(x\right):=\mathrm{C}_{f,h_{1}}\left(x\right)-\mathrm{S}_{f,\overline{h_{1}}}\left(x\right),\label{eq:Z1Z2_def}
\end{equation}
we obtain from (\ref{eq:Cfh_Sfh_coupl_simpl}) two decoupled ODE problems
\begin{equation}
Z_{+}^{\prime}\left(x\right)=f\left(x\right)\overline{Z_{+}\left(x\right)}+g_{1}\left(x\right),\hspace{1em}\hspace{1em}Z_{+}\left(0\right)=u_{1}^{0},\label{eq:Z1_pbm}
\end{equation}
\begin{equation}
Z_{-}^{\prime}\left(x\right)=-f\left(x\right)\overline{Z_{-}\left(x\right)}+g_{1}\left(x\right),\hspace{1em}\hspace{1em}Z_{-}\left(0\right)=u_{1}^{0},\label{eq:Z2_pbm}
\end{equation}
each of them are of the form (\ref{eq:u_antilin_ODE_inhom}).

\section{Some physical contexts leading to (\ref{eq:U_gen_antidiag}) and
(\ref{eq:U_gen_antidiag_inhom})\label{sec:contexts}}

\subsection{Linear Schr{\"o}dinger equation\label{subsec:cont_Schr}}

Consider the stationary linear Schr{\"o}dinger equation in $1D$,
with a potential $a\left(x\right)>0$,
\begin{equation}
u^{\prime\prime}\left(x\right)+a\left(x\right)u\left(x\right)=0,\hspace{1em}x\in\left(0,x_{0}\right).\label{eq:Schr_eq}
\end{equation}
We focus here on the initial-value problem, i.e. we supplement (\ref{eq:Schr_eq})
with the boundary conditions $u\left(0\right)=u_{0}$, $u^{\prime}\left(0\right)=u_{1}$,
but boundary-value problems on $\left(0,x_{0}\right)$, with $x_{0}$
being finite or infinite, could also be treated. We assume $a\in C^{1}\left(\left[0,x_{0}\right]\right)$. 

Introducing the vector-function 
\begin{equation}
U\left(x\right):=\left(\begin{array}{c}
u\left(x\right)\\
\frac{1}{a^{1/2}\left(x\right)}u^{\prime}\left(x\right)
\end{array}\right),\label{eq:Schr_U_vec_def}
\end{equation}
we observe that $U\left(x\right)$ satisfies
\begin{equation}
U^{\prime}\left(x\right)=\left(\begin{array}{c}
u^{\prime}\left(x\right)\\
-\frac{a^{\prime}\left(x\right)}{2a^{3/2}\left(x\right)}u^{\prime}\left(x\right)+\frac{1}{a^{1/2}\left(x\right)}u^{\prime\prime}\left(x\right)
\end{array}\right)=A\left(x\right)U\left(x\right),\label{eq:U_eq}
\end{equation}
with
\[
A\left(x\right):=\left(\begin{array}{cc}
0 & a^{1/2}\left(x\right)\\
-a^{1/2}\left(x\right) & -\frac{a^{\prime}\left(x\right)}{2a\left(x\right)}
\end{array}\right),
\]
and $U\left(0\right)=\left(u_{0},u_{1}/a^{1/2}\left(0\right)\right)^{T}$.
Here, in the second inequality of (\ref{eq:U_eq}), we used (\ref{eq:Schr_eq})
to eliminate $u^{\prime\prime}\left(x\right)$.

By writing,
\[
A\left(x\right)=a^{1/2}\left(x\right)\left(\begin{array}{cc}
0 & 1\\
-1 & 0
\end{array}\right)+\frac{a^{\prime}\left(x\right)}{2a\left(x\right)}\left(\begin{array}{cc}
0 & 0\\
0 & -1
\end{array}\right),
\]
we note that the first matrix in the right-hand side is diagonalisable
as follows
\begin{equation}
P\left(\begin{array}{cc}
0 & 1\\
-1 & 0
\end{array}\right)P^{-1}=\left(\begin{array}{cc}
i & 0\\
0 & -i
\end{array}\right),\hspace{1em}\hspace{1em}P:=\frac{1}{2^{1/2}}\left(\begin{array}{cc}
i & 1\\
1 & i
\end{array}\right),\hspace{1em}P^{-1}=\frac{1}{2^{1/2}}\left(\begin{array}{cc}
-i & 1\\
1 & -i
\end{array}\right).\label{eq:P_matr_def}
\end{equation}
Consequently, introducing $V\left(x\right):=PU\left(x\right)$, we
multiply the both sides of (\ref{eq:U_eq}) by $P$ and thus transform
it into
\begin{equation}
V^{\prime}\left(x\right)=PA\left(x\right)P^{-1}V\left(x\right)=B\left(x\right)V\left(x\right),\hspace{1em}x\in\left(0,x_{0}\right),\label{eq:V_eq}
\end{equation}
with
\[
B\left(x\right):=a^{1/2}\left(x\right)\left(\begin{array}{cc}
i & 0\\
0 & -i
\end{array}\right)-\frac{a^{\prime}\left(x\right)}{4a\left(x\right)}\left(\begin{array}{cc}
1 & -i\\
i & 1
\end{array}\right)=\left(\begin{array}{cc}
b_{0}\left(x\right) & 0\\
0 & \overline{b_{0}\left(x\right)}
\end{array}\right)-\frac{a^{\prime}\left(x\right)}{4a\left(x\right)}\left(\begin{array}{cc}
0 & -i\\
i & 0
\end{array}\right),
\]
\begin{equation}
b_{0}\left(x\right):=ia^{1/2}\left(x\right)-\frac{a^{\prime}\left(x\right)}{4a\left(x\right)},\label{eq:b0_def}
\end{equation}
and supplemented by the initial condition 
\[
V\left(0\right)=PU\left(0\right)=\frac{1}{2^{1/2}}\left(\begin{array}{c}
iu_{0}+u_{1}/a^{1/2}\left(0\right)\\
u_{0}+iu_{1}/a^{1/2}\left(0\right)
\end{array}\right).
\]

Furthermore, introducing
\[
W\left(x\right):=\left(\begin{array}{cc}
\exp\left(-\int_{0}^{x}b_{0}\left(\tau\right)\dd\tau\right) & 0\\
0 & \exp\left(-\int_{0}^{x}\overline{b_{0}\left(\tau\right)}\dd\tau\right)
\end{array}\right)V\left(x\right),
\]
we have
\[
\frac{\dd}{\dd x}\left(\begin{array}{cc}
\exp\left(-\int_{0}^{x}b_{0}\left(\tau\right)\dd\tau\right) & 0\\
0 & \exp\left(-\int_{0}^{x}\overline{b_{0}\left(\tau\right)}\dd\tau\right)
\end{array}\right)=\left(\begin{array}{cc}
-b_{0}\left(x\right)\exp\left(-\int_{0}^{x}b_{0}\left(\tau\right)\dd\tau\right) & 0\\
0 & -\overline{b_{0}\left(x\right)}\exp\left(-\int_{0}^{x}\overline{b_{0}\left(\tau\right)}\dd\tau\right)
\end{array}\right).
\]
Therefore, (\ref{eq:V_eq}) entails
\[
W^{\prime}\left(x\right)=-\frac{a^{\prime}\left(x\right)}{4a\left(x\right)}\left(\begin{array}{cc}
0 & -i\exp\left(-\int_{0}^{x}\left[b_{0}\left(\tau\right)-\overline{b_{0}\left(\tau\right)}\right]\dd\tau\right)\\
i\exp\left(\int_{0}^{x}\left[b_{0}\left(\tau\right)-\overline{b_{0}\left(\tau\right)}\right]\dd\tau\right) & 0
\end{array}\right)W\left(x\right),
\]
which, recalling (\ref{eq:b0_def}), we can rewrite as
\begin{equation}
W^{\prime}\left(x\right)=C\left(x\right)W\left(x\right),\hspace{1em}x\in\left(0,x_{0}\right),\label{eq:W_eq}
\end{equation}
with
\[
C\left(x\right):=\left(\begin{array}{cc}
0 & c_{0}\left(x\right)\\
\overline{c_{0}\left(x\right)} & 0
\end{array}\right),\hspace{1em}\hspace{1em}c_{0}\left(x\right):=\frac{ia^{\prime}\left(x\right)}{4a\left(x\right)}\exp\left(-2i\int_{0}^{x}a^{1/2}\left(\tau\right)\dd\tau\right),
\]
and the initial condition
\[
W\left(0\right)=V\left(0\right)=\frac{1}{2^{1/2}}\left(\begin{array}{c}
iu_{0}+u_{1}/a^{1/2}\left(0\right)\\
u_{0}+iu_{1}/a^{1/2}\left(0\right)
\end{array}\right).
\]

The steps described above draw from \cite{Lorenz} (see also \cite{Christ})
and provide one way to rewrite the linear Schr{\"o}dinger equation
in the form (\ref{eq:U_gen_antidiag}), but this approach is not the
only one. Alternative reduction procedures may be more cumbersome
but more beneficial in practice, depending on a final goal. For instance,
in \cite{Arnold}, the initial vectorisation of (\ref{eq:Schr_eq})
is different from (\ref{eq:Schr_U_vec_def}) yet other steps of the
transformation are ideologically similar.

\subsection{Helmholtz equation\label{subsec:cont_Helmh}}

Stationary problems for the wave propagation in heterogeneous media
are described by the Helmholtz equation whose 1$D$ version is given
by
\begin{equation}
\left(\alpha\left(x\right)u^{\prime}\left(x\right)\right)^{\prime}+\beta\left(x\right)u\left(x\right)=f\left(x\right),\hspace{1em}x\in\left(0,x_{0}\right).\label{eq:Helmh_eq}
\end{equation}
Here, $\alpha\left(x\right)$, $\beta\left(x\right)>0$ are material
parameters and $f\left(x\right)$ is the source term. As in Subsection
\ref{subsec:cont_Schr}, we suppose that (\ref{eq:Helmh_eq}) is supplemented
by the initial conditions $u\left(0\right)=u_{0}$, $u^{\prime}\left(0\right)=u_{1}$.
Furthermore, we assume that $f\left(x\right)$, $u_{0}$, $u_{1}$
are all real-valued. This assumption does not reduce generality since
(\ref{eq:Helmh_eq}) is linear with real-valued $\alpha$$\left(x\right)$,
$\beta\left(x\right)$ and hence a real-valued problem can be solved
separately for real and imaginary parts of the solution of the original
equation.

Setting 
\[
U\left(x\right):=\left(\begin{array}{c}
\alpha^{1/2}\left(x\right)\beta^{1/2}\left(x\right)u\left(x\right)\\
\alpha\left(x\right)u^{\prime}\left(x\right)
\end{array}\right),
\]
we recast (\ref{eq:Helmh_eq}) in the vector form 
\begin{equation}
U^{\prime}\left(x\right)=A\left(x\right)U\left(x\right)+F_{0}\left(x\right),\hspace{1em}x\in\left(0,x_{0}\right),\label{eq:Helm_U_eq}
\end{equation}
with 
\[
A\left(x\right):=\left(\begin{array}{cc}
\frac{\left[\alpha\left(x\right)\beta\left(x\right)\right]^{\prime}}{2\alpha\left(x\right)\beta\left(x\right)} & \left(\frac{\beta\left(x\right)}{\alpha\left(x\right)}\right)^{1/2}\\
-\left(\frac{\beta\left(x\right)}{\alpha\left(x\right)}\right)^{1/2} & 0
\end{array}\right),\hspace{1em}\hspace{1em}F_{0}\left(x\right):=\left(\begin{array}{c}
0\\
f\left(x\right)
\end{array}\right),
\]
and the initial condition
\[
U\left(0\right)=\left(\begin{array}{c}
\alpha^{1/2}\left(0\right)\beta^{1/2}\left(0\right)u_{0}\\
\alpha\left(0\right)u_{1}
\end{array}\right).
\]

We now follow the reduction steps similar to those in Subsection \ref{subsec:cont_Schr}.
We write 
\[
A\left(x\right)=\left(\frac{\beta\left(x\right)}{\alpha\left(x\right)}\right)^{1/2}\left(\begin{array}{cc}
0 & 1\\
-1 & 0
\end{array}\right)+\frac{\left[\alpha\left(x\right)\beta\left(x\right)\right]^{\prime}}{2\alpha\left(x\right)\beta\left(x\right)}\left(\begin{array}{cc}
1 & 0\\
0 & 0
\end{array}\right),
\]
and note that we can diagonalise the first matrix with the help of
the auxiliary constant matrix $P$ introduced in (\ref{eq:P_matr_def}).
Denoting $V\left(x\right):=PU\left(x\right)$, we hence have, from
(\ref{eq:Helm_U_eq}),
\begin{equation}
V^{\prime}\left(x\right)=B\left(x\right)V\left(x\right)+F_{1}\left(x\right),\label{eq:Helm_V_eq}
\end{equation}
with
\[
B\left(x\right):=\left(\begin{array}{cc}
b_{1}\left(x\right) & 0\\
0 & \overline{b_{1}\left(x\right)}
\end{array}\right)+\frac{\left[\alpha\left(x\right)\beta\left(x\right)\right]^{\prime}}{4\alpha\left(x\right)\beta\left(x\right)}\left(\begin{array}{cc}
0 & i\\
-i & 0
\end{array}\right),\hspace{1em}F_{1}\left(x\right):=PF\left(x\right)=\left(\begin{array}{c}
\frac{f\left(x\right)}{2^{1/2}}\\
i\frac{f\left(x\right)}{2^{1/2}}
\end{array}\right),
\]
\[
b_{1}\left(x\right):=i\left(\frac{\beta\left(x\right)}{\alpha\left(x\right)}\right)^{1/2}+\frac{\left[\alpha\left(x\right)\beta\left(x\right)\right]^{\prime}}{4\alpha\left(x\right)\beta\left(x\right)},
\]
and the initial condition
\[
V\left(0\right)=PU\left(0\right)=\frac{1}{2^{1/2}}\left(\begin{array}{c}
i\alpha^{1/2}\left(0\right)\beta^{1/2}\left(0\right)u_{0}+\alpha\left(0\right)u_{1}\\
\alpha^{1/2}\left(0\right)\beta^{1/2}\left(0\right)u_{0}+i\alpha\left(0\right)u_{1}
\end{array}\right).
\]

Introducing 
\[
W\left(x\right):=\left(\begin{array}{cc}
\exp\left(-\int_{0}^{x}b_{1}\left(\tau\right)\dd\tau\right) & 0\\
0 & \exp\left(-\int_{0}^{x}\overline{b_{1}\left(\tau\right)}\dd\tau\right)
\end{array}\right)V\left(x\right),
\]
equation (\ref{eq:Helm_V_eq}) transforms into
\begin{equation}
W^{\prime}\left(x\right)=C\left(x\right)W\left(x\right)+G\left(x\right),\label{eq:Helm_W_eq}
\end{equation}
where
\[
C\left(x\right):=\left(\begin{array}{cc}
0 & c_{1}\left(x\right)\\
\overline{c_{1}\left(x\right)} & 0
\end{array}\right),\hspace{1em}\hspace{1em}c_{1}\left(x\right):=\frac{i\left[\alpha\left(x\right)\beta\left(x\right)\right]^{\prime}}{4\alpha\left(x\right)\beta\left(x\right)}\exp\left(-2i\int_{0}^{x}\left(\frac{\beta\left(\tau\right)}{\alpha\left(\tau\right)}\right)^{1/2}\dd\tau\right),
\]
\[
G\left(x\right):=\left(\begin{array}{c}
g_{1}\left(x\right)\\
i\overline{g_{1}\left(x\right)}
\end{array}\right),\hspace{1em}\hspace{1em}g_{1}\left(x\right):=\frac{f\left(x\right)}{2^{1/2}}\exp\left(-\int_{0}^{x}\left[i\left(\frac{\beta\left(\tau\right)}{\alpha\left(\tau\right)}\right)^{1/2}+\frac{\left[\alpha\left(x\right)\beta\left(x\right)\right]^{\prime}}{4\alpha\left(x\right)\beta\left(x\right)}\right]\dd\tau\right),
\]
and the initial condition
\[
W\left(0\right)=\left(\begin{array}{c}
w_{1}^{0}\\
i\overline{w_{1}^{0}}
\end{array}\right),\hspace{1em}w_{1}^{0}:=\frac{1}{2^{1/2}}\left(i\alpha^{1/2}\left(0\right)\beta^{1/2}\left(0\right)u_{0}+\alpha\left(0\right)u_{1}\right).
\]
Here, in relating the first and the second components of the vector
$G\left(x\right)$, and similarly $W\left(0\right)$, we employed
the real-valuedness of $\alpha\left(x\right)$, $\beta\left(x\right)$,
$u_{0}$, $u_{1}$ that was discussed in the beginning of this Subsection.

It remains to observe that system (\ref{eq:Helm_W_eq}) is such that
the matrix $C\left(x\right)$ and the vectors $G\left(x\right)$,
$W\left(0\right)$ fit the assumptions discussed in Subsection \ref{subsec:transform_inhom}.

\subsection{Zakharov-Shabat system\label{subsec:cont_ZS}}

It is well-known that solution of a spectral problem with the linear
Schr{\"o}dinger equation appears as an intermediate step in solving
the Korteweg--de Vries (KdV) equation by using the inverse scattering
transform. Zakharov-Shabat systems play the same role in the integrability
of other nonlinear equations \cite[p.10]{Ablowitz}. In particular,
the Zakharov-Shabat system 
\begin{equation}
\partial_{x}\left(\begin{array}{c}
v_{1}\left(x,t\right)\\
v_{2}\left(x,t\right)
\end{array}\right)=\left(\begin{array}{cc}
-i\xi & q\left(x,t\right)\\
\overline{q\left(x,t\right)} & i\xi
\end{array}\right)\left(\begin{array}{c}
v_{1}\left(x,t\right)\\
v_{2}\left(x,t\right)
\end{array}\right),\label{eq:ZS_eqs}
\end{equation}
with $\xi\in\mathbb{R}$ being a spectral parameter, is a linear problem
pertinent to the integration of the defocusing cubic nonlinear Schr{\"o}dinger
(NLS) equation 
\[
i\partial_{t}q\left(x,t\right)=\partial_{x}^{2}q\left(x,t\right)-2\left|q\left(x,t\right)\right|^{2}q\left(x,t\right)
\]
subject to the initial data $q\left(x,0\right)=q_{0}\left(x\right)$.
We refer to \cite{Grebert} for more details on this matter. 

We observe that by setting 
\[
W\left(x,t\right):=\left(\begin{array}{cc}
e^{i\xi x} & 0\\
0 & e^{-i\xi x}
\end{array}\right)\left(\begin{array}{c}
v_{1}\left(x,t\right)\\
v_{2}\left(x,t\right)
\end{array}\right),
\]
Zakharov-Shabat system (\ref{eq:ZS_eqs}) immediately reduces to
\[
\partial_{x}W\left(x,t\right)=\left(\begin{array}{cc}
0 & q\left(x\right)e^{2i\xi x}\\
\overline{q\left(x\right)}e^{-2i\xi x} & 0
\end{array}\right)W\left(x,t\right),
\]
which is a system of the form (\ref{eq:U_gen_antidiag}).

\subsection{Kubelka-Munk equations\label{subsec:cont_KM}}

Kubelka-Munk equations is a simple phenomenological model for computing
reflection and transmission optical fluxes without solution of significantly
more complicated radiative transfer equations \cite{Kubelka}. Due
to their simplicity, Kubelka-Munk equations have been popular in practice
(in paper paint visibility, see e.g. \cite{Choudhury}), they have
been extensively studied from modelling viewpoint and several generalisations
have been proposed \cite{Sandoval,Yang1,Yang2,Yang3}.

We consider the following model equations
\begin{equation}
\frac{\dd}{\dd x}\left(\begin{array}{c}
F_{+}\left(x\right)\\
F_{-}\left(x\right)
\end{array}\right)=\left(\begin{array}{cc}
-K\left(x\right)-S\left(x\right) & S\left(x\right)\\
-S\left(x\right) & K\left(x\right)+S\left(x\right)
\end{array}\right)\left(\begin{array}{c}
F_{+}\left(x\right)\\
F_{-}\left(x\right)
\end{array}\right),\label{eq:KM_eqs}
\end{equation}
where $F_{+}$, $F_{-}$ are fluxes in positive and negative directions,
and $K$ and $S$ are related to absorption and scattering, respectively.
Note that, unlike in the classical model, we take here $K$, $S$
to be dependent on the optical depth $x$ rather than simply being
constants. This generalisation is expected to be useful since constant
scattering and absorption coefficients are known to be a considerable
limitation of the Kubelka-Munk model \cite[Sect. 4.5]{Choudhury}.

The procedure of reduction of (\ref{eq:KM_eqs}) to (\ref{eq:U_gen_antidiag})
is similar to that performed in Subsection \ref{subsec:cont_Schr}.
Therefore, we shall omit any detailed calculation. 

Let us write
\[
A\left(x\right):=\left(\begin{array}{cc}
-K\left(x\right)-S\left(x\right) & S\left(x\right)\\
-S\left(x\right) & K\left(x\right)+S\left(x\right)
\end{array}\right)=S\left(x\right)\left(\begin{array}{cc}
0 & 1\\
-1 & 0
\end{array}\right)-\left[K\left(x\right)+S\left(x\right)\right]\left(\begin{array}{cc}
1 & 0\\
0 & -1
\end{array}\right),
\]
and compute
\[
B\left(x\right):=PA\left(x\right)P^{-1}=S\left(x\right)\left(\begin{array}{cc}
i & 0\\
0 & -i
\end{array}\right)-\left[K\left(x\right)+S\left(x\right)\right]\left(\begin{array}{cc}
0 & i\\
-i & 0
\end{array}\right)
\]
with $P$ defined as in (\ref{eq:P_matr_def}). Multiplying the both
sides of (\ref{eq:KM_eqs}) by $P$ and introducing 
\[
V\left(x\right):=P\left(\begin{array}{c}
F_{+}\left(x\right)\\
F_{-}\left(x\right)
\end{array}\right),
\]
we obtain 
\begin{equation}
V^{\prime}\left(x\right)=B\left(x\right)V\left(x\right).\label{eq:KM_V_eq}
\end{equation}
Furthermore, setting
\[
W\left(x\right):=\left(\begin{array}{cc}
\exp\left(-i\int_{0}^{x}S\left(\tau\right)\dd\tau\right) & 0\\
0 & \exp\left(i\int_{0}^{x}S\left(\tau\right)\dd\tau\right)
\end{array}\right)V\left(x\right),
\]
\[
C\left(x\right):=\left(\begin{array}{cc}
0 & c_{2}\left(x\right)\\
\overline{c_{2}\left(x\right)} & 0
\end{array}\right),\hspace{1em}c_{2}\left(x\right):=-i\left[K\left(x\right)+S\left(x\right)\right]\exp\left(-2i\int_{0}^{x}S\left(\tau\right)\dd\tau\right),
\]
we arrive at
\begin{equation}
W^{\prime}\left(x\right)=C\left(x\right)W\left(x\right),\label{eq:KM_W_eq}
\end{equation}
which is a system of the form (\ref{eq:U_gen_antidiag}).

\section{Conclusion\label{sec:concl}}

We have introduced a new scalar differential equation of the first
order which is curious for two principal reasons. First, it is, in
some sense, the simplest nonlinear ODE (either with or without a non-homogeneous
term), with the nonlinearity being merely the complex conjugation.
Second, this equation emerges, after appropriate reduction steps,
in rather different physical contexts. Certainly, much more application
areas can be identified (e.g. telegrapher's equations or Goldstein-Taylor
model \cite{Dietert}), but already the context of the linear Schr{\"o}dinger
equation alone is a good enough motivation to further study the antilinear
ODE $u^{\prime}\left(x\right)=f\left(x\right)\overline{u\left(x\right)}$.
For example, reduction of matrix-vector manipulations to those involving
scalar quantities already provides a simplification in tedious constructions
of asymptotic-numerical methods, cf. \cite{Arnold,Korner}. Therefore,
this new reformulation yields concrete practical advantages. We believe
that theoretical aspects of the mentioned models could benefit from
it, too. This might be achievable, for instance, through newly produced
forms of the Pr{\"u}fer transformation (which is typically used for
studying Sturm-Liouville problems, see e.g. \cite[Sect. 5.2]{Pryce}).
Furthermore, it is important to identify classes of functions $f$
for which the antilinear ODE can be solved in a closed form. Here,
the Kubelka-Munk model context hints on the elementary exponential
class (note that system (\ref{eq:KM_eqs}) with constant $K$ and
$S$ is solvable explicitly). This can be generalised further since
the form of the antilinear ODE is amenable to a treatment by integral
transform methods (unlike other nonlinearities) typically compatible
with an exponential function and combinations thereof. Finally, the
form of the antilinear ODE calls for study of the possible connection
with $d$-bar problems, see e.g. \cite{Knudsen}. In this case, an
appropriate extension of the equation to the complex plane may yield
a formulation that eventually produces a closed-form solution due
to numerous constructive results on Hilbert and Riemann-Hilbert problems. 

\section*{Acknowledgements}

The author is grateful to Austrian Science Fund (FWF) for the support
through the bi-national FWF-project I3538-N32 used for his employment
at Vienna University of Technology (TU Wien). The final version of
the manuscript has benefited from valuable comments of Juliette Leblond
to whom the author is also thankful.

\end{document}